\documentclass[12pt]{article}
\usepackage{a4}
\usepackage{amssymb}
\usepackage{amstext}
\usepackage{amsmath}
\usepackage{relsize}
\usepackage{graphics}
\usepackage{graphicx}
\usepackage[T1]{fontenc}
\usepackage{makeidx}
\usepackage{theorem}
\usepackage{textcomp}

\newcommand{\IR}{{\mathbb{R}}}
\newcommand{\IN}{{\mathbb{N}}}
\newcommand{\IZ}{{\mathbb{Z}}}

\newcommand{\IP}{{\mathbb{P}}}

\newcommand{\ep}{{\varepsilon}}
\newcommand{\ph}{{\varphi}}

\newcommand{\rh}{{\varrho}}
\newcommand{\de}{{\delta}}

\newcommand{\Om}{{\Omega}}
\newcommand{\al}{{\alpha}}
\newcommand{\be}{{\beta}}
\newcommand{\ga}{{\gamma}}
\newcommand{\om}{{\omega}}
\newcommand{\La}{{\Lambda}}
\newcommand{\Ga}{{\Gamma}}
\newcommand{\De}{{\Delta}}

\newcommand{\foh}{{\mathfrak{h}}}
\newcommand{\fok}{{\mathfrak{k}}}

\newcommand{\nach}{{\rightarrow}}
\newcommand{\Nach}{{\,\rightarrow\,}}

\newcommand{\sk}{{\,|\,}}

\theoremstyle{change}
\newtheorem{satz}{Theorem}[section]

\newtheorem{lemma}[satz]{Lemma}

\newtheorem{beme}[satz]{Remarks}

\begin{document}
\title{Localization on a quantum graph with
a random potential on the edges}
\author{Pavel Exner$^{a,b}$, Mario Helm$^{c}$ and Peter Stollmann$^{c}$}

\maketitle

\begin{center}
\emph{Dedicated to Jean-Michel
Combes on the occasion of his 65th birthday}
\end{center}

\begin{quote}
{\small \em a) Nuclear Physics Institute, Academy of Sciences,
25068 \v Re\v z \\ \phantom{a) }near Prague, Czech Republic
\\
b) Doppler Institute, Czech Technical University,
B\v{r}ehov{\'a}~7, \\ \phantom{a) }11519 Prague, Czech Republic
\\
c) Technische Universit\"at Chemnitz, Fakult\"at f\"ur Mathematik,
  \\ \phantom{a) }09107 Chemnitz, Germany
\\
\phantom{a) }\texttt{exner@ujf.cas.cz},
\texttt{mario.helm@mathematik.tu-chemnitz.de},
\\ \phantom{a) }\texttt{peter.stollmann@mathematik.tu-chemnitz.de}}
\end{quote}

\begin{abstract}
\noindent We prove spectral and dynamical localization on a
cubic-lattice quantum graph with a random potential. We use
multiscale analysis and show how to obtain the necessary estimates
in analogy to the well-studied case of random Schr\"odinger
operators.
\end{abstract}

\section{Introduction}\label{c1}

Since the middle of the 1980's the mathematical approach to the
phenomenon of localization in random solids witnessed a rapid
development. One of the techniques used to prove localization is
multiscale analysis. Introduced by Fr\"ohlich and Spencer in
\cite{FS83} and further developed by von Dreifus and Klein in
\cite{DK89} for the original Anderson model on the lattice, it had
been extended to the continuum by Combes and Hislop in
\cite{CH94}. By now there is a large number of discrete and
continuum models for which localization has been established this
way, see \cite{Sto01} and for more recent advances \cite{GK01}.

On the other hand in recent years the interest also turned to
the shape of structures made of semiconductor and other materials.
In particular, quantum graph models became popular as models of
various superlattice structures. Therefore it seems natural
 to ask how one can extend the multiscale proof of localization to
such graph models. In this paper we want to give an answer for a
particular case of a special cubic lattice graph that can be
embedded in $\IR^d$, so that the known techniques work similarly
as in the ``continuum'' case. Recall that rectangular lattice
graphs also exhibit other interesting spectral
properties~\cite{Ex95}.

The embedding into $\IR^d$ provides an easy way to describe our
graph $\Ga$. Let $V(\Ga)=\IZ^d$ be the vertex set and let the set
of edges $E(\Ga)$ consist of all line segments of length one
between two neighbouring vertices in directions of the coordinate
axes. As usual we identify each edge with the interval $[0,1]$
with orientation in the sense of the increasing coordinate in
$\IR^d$. The initial and endpoint of an edge $e$ are labeled by
$\iota(e)$ and $\tau(e)$.

The embedding of $\Ga$ into $\IR^d$ allows us to define subgraphs
of $\Ga$ in terms of suitable domains in $\IR^d$. To make this
precise, we will call a bounded domain $\La\subset \IR^d$ with
piecewise smooth boundary \emph{$\Ga-$edge bounded ($\Ga$-ebdd.)}
if $\partial\La\subset E(\Ga)$ and for each edge $e\in E(\Ga)$
either $e\subset \partial \La$, or $e$ intersects $\partial \La$
at most in its endpoints. The graph $\Ga\cap\La$ arises from $\Ga$
by deleting  all the edges outside $\La$ (including those on the
boundary). For its sets of edges and vertices we write
$E(\Ga\cap\La)$ and $V(\Ga\cap\La)$, respectively.

The Hilbert space underlying our model is $L_2(\Ga):=
\bigoplus_{e\in E(\Ga)} L_2(0,1)$; in a similar way we associate
$L_2(\Ga\cap\La):= \bigoplus_{e\in E(\Ga\cap\La)} L_2(0,1)$ with
$\Ga\cap\La$. Further we need the Sobolev space of order one,
\begin{eqnarray*}
    W_2^1(\Ga)&:=& \{f\in \bigoplus_{e\in E} W_2^1(0,1)
    \sk f \text{ continuous at all vertices }
    v\in V,\\ && \|f\|^2_{W_2^1(\Ga)}:=
    \sum_{e\in E(\Ga)} \|f_e\|^2_{W_2^1(0,1)} <\infty \}
\end{eqnarray*}
with the obvious notation and terminology for edge components of
$f$, and its analogue $W_2^1(\Ga\cap\La)$.

We  can now define the random Schr\"odinger operator
$H(\om)$  for $\omega\in\Omega:=[q_-,q_+]^{E}$
 via their associated forms,
\begin{eqnarray*}
    D(\foh_\om)&=& W_2^1(\Ga),  \\
    \foh_\om(f,g)&=& \sum_{e\in E(\Ga)} \,[
    (f_e'\sk g_e')_{L_2(0,1)} + (\om_e\cdot f_e \sk
    g_e)_{L_2(0,1)}]
\end{eqnarray*}
These self-adjoint operators correspond to the
differential expression $-f_e''+\om_e\cdot f_e$ on the edges,
together with the free (often called Kirchhoff) boundary
conditions at the inner vertices, i.e.
\begin{equation*}\label{g006}
    \sum_{\iota(e)=v} f_e'(0) - \sum_{\tau(e)=v} f_e'(0)
    = 0 \qquad(\forall v\in V\cap \Lambda),
\end{equation*}
The coupling constants $\om_e$ carry the random structure. They
are picked independently for different edges with a probability
measure $\mu$ on $\IR$ with supp$\,\mu= [q_-,q_+]$. For technical
reasons we have to assume that $\mu$ is H\"older continuous with
H\"older exponent $\al$ and further that $\mu$ satisfies the
following assumption: there exists $\tau>\frac{d}{2}$ such that
for $h$ small
\begin{equation}\label{g002}
    \mu([q_-,q_{-}+h])\leq h^{\tau}.
\end{equation}
This \emph{single site measure} $\mu$ defines a probability
$\mathbb{P} :=\bigotimes_{e\in
E}\:\mu$ on $\Om$.

We will also need restrictions $H^N_\La(\om)$ for an ebbd open
$\La$ defined via the form
\begin{eqnarray*}\label{g008}
   D(\foh_{\La,\om}^N)&=&W_2^1(\Ga\cap\La) \nonumber \\
   \foh_{\La,\om}^N(f,g)&=& \sum_{e\in E(\Ga\cap\La)} \,[
   (f_e'\sk g_e')_{L_2(0,1)} + (\om_e\cdot f_e \sk
   g_e)],
\end{eqnarray*}
which corresponds to Neumann boundary conditions at the boundary
vertices $v\in V\cap \partial\Lambda$ -- cf.~\cite{Ku04}.


\section{The main results and the idea of their proof}\label{c2}

Our family of random Schr\"odinger operators exhibits
deterministic spectrum, i.e. there exists a closed subset
$\Sigma\subset \IR$ s.t. $\sigma(H(\om))=\Sigma$ almost surely.
This is a standard result from the theory of random operators --
see, e.g., \cite{CL90} -- and comes from fundamental properties of
our construction, especially the ergodicity w.r.t. lattice
translations. To locate the deterministic spectrum we can consider
the free operator $H_0$ (the one with $V=0$) and use some results
that relate the spectrum of $H_0$ to the spectrum of its
transition operator, the  Laplacian on $\IZ^d$ -- see, e.g.,
\cite{Ex97, Cat97}. In this way we get $\sigma(H_0)= [0,\infty)$
and hence again by standard theory $\Sigma=[q_-,\infty)$.

Our first claim is that in some neighborhood of $\inf\Sigma=q_-$
the operators exhibit pure point spectrum with exponentially
decaying eigenfunctions almost surely:

\begin{satz}[Spectral/Anderson localization]\label{s020}
There is an $\ep>0$ such that the spectrum of $H(\om)$ in
$[q_-,q_-+\ep_0]$ is pure point for a.e. $\om\in\Om$. Furthermore,
there exists a $\ga>0$ and for each eigenfunction $u$ associated
to an energy in this interval a constant $C_u$ such that
\begin{equation*}
   \|\chi_{\La_1(x)}u\| \leq C_u\cdot
   \exp[-\gamma \,d(x,0)] \qquad (x\in\Ga),
\end{equation*}
where $\La_1(x)$ is the intersection of $\Ga$ with the unit cube
centered at $x\in\IZ^d$.
\end{satz}

The assertion of the preceding theorem is sometimes called
Anderson localization or spectral localization (see \cite{RJLS95}
for a discussion of different concepts of localization). An
alternative and stronger concept is  dynamical localization, see
\cite{GB98, DSt01} and \cite{GK01} for more recent developments.
In the context of our model the following result is valid.

\begin{satz}[Strong dynamical localization]\label{s026}
Let $p>2(2\tau-d)$ where $\tau$ refers to (\ref{g002}). Then there
exists an $\ep>0$ such that for $K\subset\Ga$ compact, each
interval $I\subset [E_{0},E_{0}+\ep]$ and $\eta\in
L_{\infty}(\IR)$ with $\text{supp } \eta \subset I$ we have
\begin{equation*}
   \mathbb{E}\{\||X|^{p}\eta(H(\omega))\chi_{K}\| \} <  \infty,
\end{equation*}
which in particular means that
\begin{equation*}
   \mathbb{E}\{\sup_{t>0} \||X|^{p}e^{-itH(\omega)}
   P_{I}(H(\omega))\chi_{K}\| \} < \infty.
\end{equation*}
\end{satz}
Both results will be proved by a multiscale induction as presented
in detail in \cite{Sto01}. As the framework introduced there is
general enough to include our case it will be sufficient to
establish the necessary model-dependent estimates that are to be
plugged into the multiscale machinery.

For the readers convenience we will now briefly describe the idea
behind the multiscale induction. The basic property one proves by
induction is an exponential decay estimate for the kernel of the
resolvent of $H^N_{\La(L)}(\om)$. More precisely, it is shown that
with high probability (depending on $L$) the resolvent of
$H^N_{\La(L)}(\om)$ shows exponential off-diagonal decay.

Note that, outside the spectrum of a Schr\"odinger operator, such
an exponential decay estimate is just the content of the
celebrated \emph{Combes-Thomas estimate}. We will make clear that
an analogue holds for quantum graphs as well. Actually, this kind
of argument will give the starting point of our induction
procedure, the \emph{initial length scale estimate}. More
precisely, the assumption (1) on the tail of the single site
measure implies that energies near $\inf\Sigma$ are in the
resolvent set of $H^N_{\La(L)}(\om)$ with high probability for any
given $L$. However, keeping an interval near $\inf\Sigma$ fixed
and letting $L$ tend to infinity, the interval will be filled with
eigenvalues of the box Hamiltonian. Therefore the sought property,
the exponential decay, must be deduced by a more clever argument.
One important ingredient is the relation between resolvents of
different nested boxes, cast in the form of a \emph{geometric
resolvent identity}. This will allow to conclude exponential decay
on a large box, knowing exponential decay on smaller sub-boxes. In
this induction step, from length $L$ one proceeds to $L^\al$ with
suitable $\al>1$. A very important a priori information is
necessary, the so-called \emph{Wegner estimate}. Putting these
estimates together as in \cite{Sto01} one arrives at the desired
exponential decay estimates for larger and larger boxes. To
conclude, finally, that the operators $H(\om)$ exhibit pure point
spectrum almost surely, we need to know that the spectrum is
indeed determined by \emph{generalized eigenfunctions}. In the
next section we show how to obtain these steps.

\begin{beme}
{\rm (a) Our results can easily be extended to certain other
cases, for instance, to a ``rhombic'' lattice, where the present
method would work after adjusting constants appearing in the
equivalence
between the Euclidean and the intrinsic metric. \\
(b) The results could be also extended to potentials, which are
only relatively bounded, for instance, one can consider suitable
$L_p(0,1)-$functions with a positive lower bound as ``single
edge'' potentials, following \cite{Sto01} and numerous other
papers; we did not take this path and treated characteristic
functions as random potentials here exclusively for the sake of
simplicity. \\
(c) In a different direction, results are available for certain
random quantum graphs, namely for random trees with random edge
lengths; see the recent work in \cite{ASW, HP}.  }

\end{beme}


\section{The proofs}

\subsection{A Combes-Thomas estimate}\label{c7}

The statements of this section will show how to obtain
``exponential decay of the local resolvent'' outside the spectrum.
The results go back to the celebrated paper \cite{CT} and its
improvement in \cite{BCH97}.

\begin{satz}[Combes-Thomas estimate]\label{s010}
Let $R>0$. There exist constants $c_1=c_1(q_-,q_+, R)$,
$c_2=c_2(q_-,q_+, R)$, s.t. from the assumptions
\begin{itemize}
    \item [(i)] $\La\subset \IR^d$  $\Ga-$ebdd. box, $A,B \subset \La\,$
    $\Ga-$ebdd., dist$(A,B)=:\de\geq 1$,
    \item [(ii)] $(r,s)\subset \rh(H_\La^N)\cap(-R,R)\,$, $E\in
    (r,s)$, $\eta:=$ dist$(E,\, (r,s)^c)>0$
\end{itemize}
it follows that
\begin{equation*}\label{g038}
    \| \chi_A (H_\La^N -E)^{-1}\chi_B\| \leq c_1 \cdot \eta^{-1}\cdot
    e^{-c_2\sqrt{\eta(s-r)}\,\de}.
\end{equation*}

\end{satz}
\textbf{Proof: } Let $ w: \La\Nach\IR$ be defined as $w(x):=
\text{dist}(x,B).$ By triangle inequality
\begin{equation*}
    |w(y)-w(x)| \leq |x-y|,
\end{equation*}
so that $\|\nabla w\|_\infty\leq 1,$ and this in turn implies
$\|w'\|_\infty\leq 1$ for the restriction to the graph.
Furthermore, the functions $\psi(x)= e^{-w(x)}$ and $\ph(x)=
e^{w(x)}$ are uniformly Lipschitz continuous on all edges because
\begin{eqnarray*}
    |e^{w(y)}-e^{w(x)}| &\leq& \sup_{\xi\in \Ga\cap \La} |
    (\exp\circ w)'(\xi)| \cdot |y-x|  \\
    &\leq& \sup_{\xi\in \Ga\cap \La} |\exp(w(\xi))||w'(\xi)| \cdot |y-x|.
\end{eqnarray*}
Hence for each $u\in D(\foh)$ also the functions $\psi u, \ph u$
belong to $D(\foh)$, which means that
\begin{equation*}
    \foh_\be(u,v) := \foh(e^{-\be w}u, e^{\be w} v)
\end{equation*}
is well defined for all $u,v \in D(\foh)$. By the product rule we
have the relation
\begin{eqnarray*}
     \foh_\be(u,v)
     &=& (e^{-\be w} u' \sk e^{\be w} v') -\be ((e^{-\be w}uw'\sk  e^{\be w} v') \\
     && - \be^2 ((e^{-\be w}uw' \sk e^{\be w} vw') + \be((e^{-\be w}u'\sk e^{\be w}
     vw')+ (Vu \sk v)\\
     &=& \foh(u,v) - \be\underbrace{[(uw'\sk v')-(u\sk vw')]}_{(\ast)} - \be^2(w'^2 u \sk
     v).
\end{eqnarray*}
Referring to the term $(\ast)$ above we define the symmetric form
\begin{equation*}
    \fok (u,v) := i [(uw'\sk v')-(u\sk vw')].
\end{equation*}
Using $1 \geq m:= w'^2\geq 0$ one can write
\begin{eqnarray*}
    \foh_\be(u,v) &=&  \tilde{\foh}(u,v) + i\be \fok (u,v), \quad \text{where} \\
    \tilde{\foh}(u,v) &=& \foh(u,v) - \be^2(mu\sk v).
\end{eqnarray*}
Next we are going to show that $\foh_\be$ is sectorial. From
$\|w'\|_\infty \leq 1$ one gets
\begin{eqnarray}\label{g043}
   \fok (u) \leq 2 \|u\|\,\|u'\| \leq \|u'\|^2 + \|u\|^2 .
\end{eqnarray}
On the other hand, consider the operator $\tilde{H}$ associated
with $\tilde{\foh}$ and $C=C(R)$,\, $C\geq \be^2+1, \,C\geq 1-r$m
for which we have
\begin{eqnarray}\label{g044}
    \|(\tilde{H}+C)^\frac{1}{2} u \|^2 &\geq& \|(\tilde{H}+\be^2+1)^\frac{1}{2} u \|^2\nonumber \\
    &=& \|u'\|^2 + ([V+ \be^2\underbrace{(1-m)}_{\geq 0} +1] u \sk u) \nonumber \\
    &\geq& \|u'\|^2 +\|u\|^2.
\end{eqnarray}
It follows from (\ref{g043}) and (\ref{g044}) that
\begin{equation}\label{g046}
    |\fok (u)| \leq \|(\tilde{H}+C)^\frac{1}{2} u \|^2 = (\tilde{\foh}+
    C)(u),
\end{equation}
hence $\foh_\be = \tilde{\foh} + i\be \fok$ is sectorial and there
exists an associated sectorial operator  $H_\be$ -- see, e.g.,
\cite{Kato76}.

In the next step we are going to show the existence of a
bounded operator $S$ on $L_2(\Ga\cap\La)$, $\|S\|\leq 1$, s.t.
\begin{equation*}
    \fok(u,v) = ( S(\tilde{H}+C)^\frac{1}{2} u \sk (\tilde{H}+C)^\frac{1}{2}
    v) \qquad(\forall u,v \in D(\foh)).
\end{equation*}
Let thus $D(\foh)$  be equipped with the scalar product
$(\tilde{h}+C)(\cdot,\cdot) $. By the Riesz representation theorem
there exists a bounded operator $K$ on $D(\foh)$ with
\begin{equation*}
    \fok (u,v) = (\tilde{\foh}+C)(Ku,v)
\end{equation*}
and by (\ref{g046}) we have $\|K\|\leq 1$. Put
\begin{equation*}
    S:= (\tilde{H}+C)^\frac{1}{2} K (\tilde{H}+C)^{-\frac{1}{2}}:
    L_2(\Ga\cap\La)\Nach L_2(\Ga\cap\La).
\end{equation*}
As $(\tilde{H}+C)^\frac{1}{2}: D(\foh)\Nach L_2(\Ga\cap\La)$ and
$(\tilde{H}+C)^{-\frac{1}{2}}: L_2(\Ga\cap\La) \Nach D(\foh)$ are
unitary, we have $\|S\| = \|K\| \leq 1$, and for $u,v \in D(\foh)$
we get the desired relation
\begin{eqnarray*}
    ((\tilde{H}+C)^\frac{1}{2}u \sk (\tilde{H}+C)^\frac{1}{2}v)
    &=& ((\tilde{H}+C)^\frac{1}{2}Ku \sk (\tilde{H}+C)^\frac{1}{2}v) \\
    &=& \fok (u,v).
\end{eqnarray*}
Now we have to investigate invertibility of $H_\be-E$ for $E\in
(r,s)$ in dependence on $\be$. Here we can use the proof of
\cite{Sto01} (which in turn uses Lemma 3.1. from \cite{BCH97})
word by word, so we present just the result: let
\begin{equation*}
     \be_1:= \min \left\{\be_0,\, \frac{1}{R+C}
     \sqrt{\frac{1}{32}\,\eta\, (s-r)} \right\},
\end{equation*}
then for $|\be|\leq \be_1$ the operator $T+i\be S$ is invertible
with
\begin{equation}\label{g050}
    \|(T+ i\be S)^{-1}\| \leq 4 \,\frac{R+C}{\eta}.
\end{equation}
Next we will find a connection between $T+ i\be S$ and $H_\be-E$
which shows that for $|\be|\leq \be_1$ the operator $H_\be-E$ is
invertible too, namely
\begin{equation}\label{g052}
     (H_\be-E)^{-1} = (\tilde{H}+C)^{-\frac{1}{2}}(T+i\be S)^{-1}
     (\tilde{H}+C)^{-\frac{1}{2}}.
\end{equation}
Let $f\in L_{2}(\Ga\cap\La)$, then
   \begin{equation*}
     u:= (\tilde{H}+C)^{-\frac{1}{2}}(T+i\beta S)^{-1}
     (\tilde{H}+C)^{-\frac{1}{2}} f \in D(\foh)
   \end{equation*}
holds, since $ (\tilde{H}+C)^{-\frac{1}{2}}$ maps $L_{2}(\Lambda)$
to $D(\foh)$. Using the definitions of $T, S$ and $u$ we can
calculate for $v\in D(\foh)$ the expression
   \begin{eqnarray*}
     (\mathfrak{h}_{\beta}-E)(u,v) &=& (\widetilde{\mathfrak{h}}-E)(u,v)
     + i\beta\,\mathfrak{k}(u,v)\\
     &=& (T(\widetilde{H}+C)^{\frac{1}{2}}u|(\widetilde{H}+C)^{\frac{1}{2}}v)+
     i\beta\,
     (S(\widetilde{H}+C)^{\frac{1}{2}}u|(\widetilde{H}+C)^{\frac{1}{2}}v)\\
     &=& ((T+i\beta S)(\widetilde{H}+C)^{\frac{1}{2}}
     u|(\widetilde{H}+C)^{\frac{1}{2}}v)\\
     &=& ((\widetilde{H}+C)^{-\frac{1}{2}} f|(\widetilde{H}+C)^{\frac{1}{2}}v)\\
     &=& (f|v).
   \end{eqnarray*}
Consequently, we have $u\in D(H_{\beta}-E)$ and $(H_{\beta}-E)u =
f$, so (\ref{g052}) follows, and by (\ref{g050}) we get
\begin{equation}\label{g054}
     \|(H_\be-E)^{-1}\| \leq 4\,\frac{R+C}{\eta}.
\end{equation}
A straightforward calculation now shows that
\begin{equation*}
     (H_\be-E)^{-1} f = e^{\be w}(H-E)^{-1} e^{-\be w} f,
\end{equation*}
and therefore
\begin{equation}\label{g056}
     \|\chi_A (H-E)^{-1} \chi_B\| \leq \|\chi_A e^{-\be
     w}\|_{\infty}\cdot \|(H_{\be}-E)^{-1}\|\cdot \|e^{\be
     w}\chi_B\|_{\infty}.
\end{equation}
Putting $\be:=\frac{1}{2}\be_1$ we analyze the factors on the
right-hand side. By $w|_B=0$ one has $\|e^{\be
w}\chi_B\|_{\infty}$ $\leq 1$. The second factor is controlled by
(\ref{g054}), and furthermore, by definition of $\be_1$ there is a
constant $c_2=c_2(R)$ s.t.
\begin{equation*}
    \be\geq c_2(R) \cdot \sqrt{\eta(s-r)}.
\end{equation*}
By assumption, $w(x)=\mathrm{dist}(x,B)\geq \de$ for all $x\in A$,
i.e.
\begin{equation*}
     \|\chi_A e^{-\be w}\|_{\infty} \leq e^{-\be\cdot
     \de} \leq \exp (-c_{2}(R)\cdot\sqrt{\eta(s-r)}\cdot \delta) .
\end{equation*}
Combining  this argument with (\ref{g056}) we get finally the
result,
\begin{equation*}
    \|\chi_A (H-E)^{-1} \chi_B\| \leq c_1(R)\cdot
    \eta^{-1}\cdot \exp (-c_{2}(R)\cdot\sqrt{\eta(s-r)}\cdot
    \delta).
\end{equation*}

\subsection{The initial length scale estimate}\label{c5}

The initial length scale estimate tells us something about the
probability that an eigenvalue of the box hamiltonian is found
inside a suitable interval. Specifically, we take an interval
centered at the lower bound $q_-$ of the deterministic spectrum
and  we suppose that its length depends on the size $l$ of the
box. The estimate we are interested in will only hold for lengths
larger than some initial value $l^\ast$.

\begin{satz}[Initial length scale estimate]\label{s004}
For each $\xi\in (0,2\tau\!-\!d)$ there exist
$\beta=\beta(\tau,\xi)\in (0,2)$ and $l^{\ast}=l^{\ast}(\tau,\xi)$
such that
   \begin{equation}\label{g024}
     \IP \{\text{dist}(\sigma(H_\La^N(\omega)), q_-) \leq l^{\beta
     -2}\}\,\leq\, l^{-\xi}
   \end{equation}
holds for all $\Ga-$ebdd. boxes $\La=\La_{l}(0)\,$ with $l\geq
l^{\ast}$.
\end{satz}
\emph{Proof: } Let
\begin{equation*}
    \Om_{l,h} := \{\omega \in\Omega \sk \, q_e(\omega)
    \geq q_- +h \text{ for all } e\in
    E(\Ga\cap \La\}.
\end{equation*}
By the min-max principle we infer that for $\om\in \Om_{l,h}$
\begin{equation*}
    E_0(H_\La^N) \geq E_0((-\De + q_- + h)_\La^N) = q_- +h,
\end{equation*}
where $E_0$ is the lowest eigenvalue of the respective operator.
Using assumption (\ref{g002}) the probability of $\Om_{l,h}$ can
be estimated by
\begin{eqnarray*}
   \IP(\Om_{l,h}) &\geq& 1- \sharp E(\Ga\cap\La)\cdot \mu([q_-,q_- + h]) \\
                  &\geq& 1- d\cdot|\La|\cdot h^\tau.
\end{eqnarray*}
Let $\xi\in (0,2\tau\!-\!d)$. Then it is always possible to choose
$\beta\in (0,2)$ such that
   \begin{equation*}
     \xi < \tau(2-\beta)-d,
   \end{equation*}
and inserting $h:= l^{\beta -2}$ we get for $l$ large
\begin{eqnarray*}
   \IP(\Om_{l,h}) &=& 1- d\, |\La| \,  l^{\tau(\beta -2)}  \\
   &=& 1- \underbrace{d\,l^{\xi-\tau(2-\beta)+d}}_{\leq 1 \text{ for  $l$ large}}\, l^{-\xi} \\
   &\geq& 1-l^{-\xi}.
\end{eqnarray*}
\subsection{The geometric resolvent inequality}\label{c6}

As we mentioned above, in the multiscale induction step one has to
deal with restrictions of a Schr\"odinger operator to nested cubes
on different length scales. Consequently, we need a tool that
relates the resolvents of such restrictions. The first step on
this way is the following lemma, called geometric resolvent
equality.

\begin{lemma} [Geometric resolvent equality]\label{s006}
   Let $\La\subset\La' \subset\IR^d$ be some open $\Ga-$ebdd. boxes,
   $H_{\La}$ and $H_{\La'}$ the respective realizations
   of our model operator with Neumann b.c. Let $\psi\in
   \{f|_{\Ga\cap\La}\sk f\in C_{c}^{1}(\La)\}$ be real-valued.
   Then we have for each $z\in \varrho(H_{\Lambda})\cap
   \varrho(H_{\Lambda'})$ the relation
   \begin{equation*}\label{g026}
      R_{\La}\psi = \psi\, R_{\La'}
      + R_{\La}\,[\psi'\cdot D + D\,\psi']\, R_{\La'},
   \end{equation*}
   where we have denoted $R_{\Lambda}:=(H_{\Lambda}-z)^{-1},\,
   R_{\Lambda'}:=(H_{\Lambda'}-z)^{-1}$, $D$ is the first
   derivative, and all the terms are interpreted as operators
   on $L_{2}(\Ga\cap\La')$.
\end{lemma}
\emph{Proof: } We regard $L_{2}(\Ga\cap\La)$ as a subspace of
$L_{2}(\Ga\cap\La')$. In terms of the associated forms the
assertion then reads as follows:
\begin{eqnarray*}
    (\foh_\La-z) (\psi\, R_{\La'} + R_{\La}\,[\psi'\cdot D + D\,\psi']\,
    R_{\La'})g ,\, w) &=& (\psi g \sk w) \\
    &&(\forall g \in L_2(\Ga\cap\La), w \in D(\foh));
\end{eqnarray*}
notice that in this case the first argument at the left-hand side,
which we denote as $u$, belongs to $D(H_\La)$ and
$(H-z)u=\psi\cdot g$.

In the first step we have to show that $u\in D(\foh)$ holds. By
the product rule, $\psi_e\, (R_{\La'}g)_e \,\in\, W_2^1(0,1)$ for
all $e\in E(\Ga\cap\La)$. The continuity of $\psi\, R_{\La'} g$ at
the inner vertices of $\La'$ is clear, so the first term is
controlled. Further we find $\psi' D R_{\La'}: \, L_2(\Ga\cap\La)
\Nach L_2(\Ga\cap\La)$, i.e.
\begin{equation*}
    R_\La \psi' D R_{\La'}\, g \,\in\, D(\foh).
\end{equation*}
For the analysis of the third term one has
   \begin{equation*}
      \psi' R_{\La'}\,g \in L_{2}(\Ga\cap \La).
   \end{equation*}
Now $R_\La \, D: W_2^1(\Ga\cap\La)\Nach D(\foh_\La)$ extends to a
bounded operator from $L_2(\Ga\cap\La)$ to $D(\foh_\La)$. Indeed,
we can always choose $z$ small enough, in which case
\begin{equation*}
   R((H_{\La}-z)^{-\frac{1}{2}})=D((H_{\La}-z)^{\frac{1}{2}})=
   D(\foh_\La)\subset W_2^1(\Ga\cap\La).
\end{equation*}
For $v\in D(\foh_\La)$ we have
\begin{eqnarray*}
   \|v'\|_{L_2(\Ga\cap\La)}^2 &=& \foh(v)-
   \sum_{e\in E(\Ga\cap\La)} V_e \int_0^1 \,v_e^2(x) dx  \\
   &\leq& \|v\|_{D(\foh_\La)},
\end{eqnarray*}
i.e. $D \,(H_{\La}-z)^{-\frac{1}{2}}$ is bounded on
$L_2(\Ga\cap\La)$. Thus for $\ph\in W_2^1(\Ga\cap\La)$ and $f\in
L_2(\Ga\cap\La)$ we get
\begin{eqnarray*}
   |((H_{\La}-z)^{\frac{1}{2}}\ph' \sk f)| &=& |(\ph|\ D\,
   (H_{\La}-z)^{-\frac{1}{2}}f)| \\
   &\leq& c \cdot \|\ph\|_{L_2(\Ga\cap\La)}\cdot
   \|f\|_{L_2(\Ga\cap\La)},
\end{eqnarray*}
and from here finally the boundedness of the map
\begin{equation*}
   R_{\La} D =
   (H_{\La}-z)^{-\frac{1}{2}}(H_{\La}-z)^{-\frac{1}{2}}D :\,
   L_2(\Ga\cap\La) \Nach L_2(\Ga\cap\La) \Nach  D(\foh_\La).
\end{equation*}

The next step is to control the behavior of some functions at the
inner vertices. For a fixed inner vertex of $\Ga\cap\La$ let
$e_{k,\text{in}}$ and $e_{k,\text{out}}$ be the in- and outcoming
edges, respectively, parallel to the $k-$th coordinate axis, and
let $\partial_k \psi(v)$ be the $k-$th partial derivative of the
$C_c^1(\La)-$continuation of $\psi$. Then
\begin{eqnarray*}
   (D \psi' R_{\La'}g\sk w)_{L_2(\Ga\cap\La)} &=& \sum_{e\in E(\Ga\cap\La)}
   (D \psi'_e R_{\La'}g_e\sk w_e)_{L_2(0,1)} \hspace{3cm}
\end{eqnarray*}\vspace{-6mm}
\begin{eqnarray}\label{g030}
   \hspace{1,5cm}&=&   \sum_{e\in E(\Ga\cap\La)}
   \{(-\psi'_e R_{\La'}g_e\sk w'_e)_{L_2(0,1)}+
   \psi'_e R_{\La'}g_e w_e\,|_0^1\} \nonumber \nonumber \\
   &=&  -(\psi' R_{\La'}g \sk w')_{L_2(\Ga\cap\La)} \nonumber\\
   &&+ \sum_{v \text{ inn. vertex}}\, \sum_{k=1}^d \partial_k \psi(v)
   \{\underbrace{(R_{\La'} g w)_{e_{k, \text{in}}}(1)
   -(R_{\La'} g w)_{e_{k, \text{out}}}(0)}_{= 0
   \text{ by continuity at inner vertices}} \} \nonumber \\
   &=&-(\psi' R_{\La'}g \sk w')_{L_2(\Ga\cap\La)}.
\end{eqnarray}
The following calculation now finishes the proof:
\begin{eqnarray*}
   (\foh_\La-z)(u,w) &=& (\foh_\La-z)(\psi\,
   R_{\La'} g,w)+( (\psi'\cdot D R_{\La'}
   + D\,\psi'R_{\La'})g \sk w)\\
   &\stackrel{(\ref{g030})}{=}&
   ((\psi\, R_{\La'} g)'\sk w')+ ((V-z)\psi\, R_{\La'} g \sk w)\\
   && + (\psi' (R_{\La'}g)' \sk w)-(\psi' R_{\La'}g \sk w') \\
   &=& (\psi'\, R_{\La'} g\sk w')+ (\psi (R_{\La'} g)'\sk w')
   + (\psi' (R_{\La'}g)' \sk w)\\
   && + ((V-z)\psi\, R_{\La'} g \sk w) -(\psi' R_{\La'}g \sk w')  \\
   &\stackrel{\psi \text{ real val.}}{=}&  ((R_{\La'} g)'\sk (\psi w)')
   + ((V-z)\, R_{\La'} g \sk \psi w)\\
   &=&  (\foh_{\La'}-z)(R_{\La'}g,\psi w)\\
   &=&  (g\sk \psi w)\\
   &=&  (\psi g\sk w).
\end{eqnarray*}

\bigskip

We will next prove another preparatory lemma after which we will
be ready to state the main theorem of this section.
\begin{lemma}\label{l006}
Let $\tilde{\Om}\subset\Om\subset\IR^d$ be a $\Ga-$ebdd. domains,
$\text{dist}(\partial\tilde{\Om},\partial\Om)  >0$,\, $E\in \IR$
and $g\in L_2(\Ga\cap\Om)$. Then there exists $C=C(q_-, q_+, E)$
s.t. for all $u\in W_2^1(\Ga\cap\Om)$ with
\begin{equation*}
    (u'\sk \ph')_{L_2(\Ga\cap\Om)}+ (Vu \sk \ph)_{L_2(\Ga\cap\Om)}
    = (g\sk \ph)_{L_2(\Ga\cap\Om)}
    \qquad (\forall \ph\in W_{2,0}^1(\Ga\cap\Om))
\end{equation*}
it holds that
\begin{equation*}
     \|u'\|_{L_2(\Ga\cap \tilde{\Om})}\leq
     C(\|u\|_{L_2(\Ga\cap\Om)}+ \|g\|_{L_2(\Ga\cap\Om)}).
\end{equation*}
\end{lemma}
\textbf{Proof: } By construction, dist$(\partial\Om,
\partial\tilde{\Om}) \geq 1$, hence there exists a vector
\begin{equation*}
    \psi\in \{f|_{\Ga\cap\Om}\sk f\in C_c(\Om),\,
    \text{supp\,}f' \subset \{x\in\Om \sk \text{dist}(x,\tilde{\Om})<1\}   \}
\end{equation*}
with $0\leq \psi\leq 1,\, \psi|_{\Ga\cap\tilde{\Om}}=1$ and
$\|\psi'\|_\infty\leq \tilde{C}(d)$. Let $w:=u\psi^2$, then $w\in
W_{2,0}^1(\Ga\cap\Om)$, and by product rule we find
\begin{equation*}
    (u'\sk w')_{L_2(\Ga\cap\Om)} = (\psi u'\sk \psi u')+ 2(\psi u' \sk u\psi').
\end{equation*}
Using $\tilde{V}:=V-E$ and support properties of the functions
involved we get
\begin{eqnarray*}
   \|\psi u'\|^2 &=&  (u'\sk w') - 2(\psi u' \sk u\psi')\\
   &=& (g\sk w)-(\tilde{V} u \sk w)- 2(\psi u' \sk u\psi') \\
   &\leq& \|g\|\,\|u\| + |(\tilde{V}\psi u\sk \psi u)|
   + 2\,\|\psi u'\|\,\|u\|\,\|\psi'\|_\infty  \\
   &\leq& \|g\|\,\|u\| + \hat{C}(q_-, q_+, E) \|u\|^2 + 2\tilde{C}
   \,\|\psi u'\|\,\|u\|.
\end{eqnarray*}
We consider the latter as a quadratic inequality in $\|\psi u'\|$,
and find after some simple manipulations, that it can only be
fulfilled for
\begin{eqnarray*}
   \|\psi u'\|&\leq&  \sqrt{\tilde{C}^2+ \hat{C}}\,\|u\|
   + \frac{1}{2\sqrt{\tilde{C}^2+
    \hat{C}}}\,\|g\| \\
   &=& C(q_-, q_+, E)(\|u\|+\|g\|)
\end{eqnarray*}
By $\psi|_{\Ga\cap\tilde{\Om}} =1$ the assertion follows.
$\;\square$

\medskip

Before we come to the main point we introduce some notation. A
$\Ga-$ebdd. box $\La=\La_L(x)$ is called \emph{suitable}, if $x\in
\IZ^d$, $L\in 6\IN\setminus 12\IN$ and $ L\geq 42$. For such boxes
we define
\begin{eqnarray*}
   \La_\text{int}(x) = \La_{L, \text{int}(x)} &:=& \La_{L/3}(x), \\
   \La_\text{out}(x) = \La_{L, \text{out}(x)}&:=& \La_L(x)\setminus \La_{L-12}(x)
\end{eqnarray*}
and write for the respective characteristic functions on the
graph:
\begin{equation*}
   \chi_\La^\text{int} = \chi_{\La_L(x)}^\text{int}:= \chi_{\Ga \cap
   \La^\text{int}(x)}^\text{int},\quad
   \chi_\La^\text{out} = \chi_{\La_L(x)}^\text{out}:= \chi_{\Ga \cap
   \La^\text{int}(x)}^\text{out}.
\end{equation*}
In general the symbol $\chi_A$ for a $\Ga-$ebdd. domain is to be
understood as $\chi_{\Ga\cap A}$.

\begin{satz}[Geometric resolvent inequality] \label{s008}
Let $\La\subset \La' \subset \IR^d$ be suitable $\Ga-$ebdd. boxes.
Let further $A\subset \La^\text{int}$ and $  B\subset
\La'\setminus\La$ be $\Ga-$ebdd. domains, $I_0\subset \IR$ bounded
and $E\in I_0$. Then there exists
$C_{\text{geom}}=C_{\text{geom}}(q_-,q_+, E)$ s.t.
\begin{equation*}\label{g032}
    \|\chi_B R_{\La'}(E)\chi_A\| \leq C_{\text{geom}}\cdot
    \|\chi_B R_{\La'}(E)\chi_\La^{\text{out}}\|\,\|\chi_\La^{\text{out}}
    R_\La(E)\chi_A\|.
\end{equation*}
\end{satz}
\textbf{Proof: } Let $x\in \IZ^d$ be the center of $\La$. We
choose $\ph\in \{f|_{\Ga\cap \La}\sk f\in C_c^\infty (\La)\}$
real-valued with $\mathrm{supp\,}f \subset \La_{L-4}(x)$ s.t.
$\ph=1$ on $\La_{L-8}(x)$. This can be certainly achieved, with
$\|\ph'\|_\infty$ bounded independent on $\La$.

Let $\Om:=\text{int}\, \La^\text{out}$, i.e.
$\text{dist}\,(\partial\Om,\, \text{supp}\,\ph')\geq 2.$ By the
geometric resolvent equality (Lemma \ref{s006}) we have
\begin{eqnarray*}
   \|\chi_B R_{\La'} \chi_A\| &=&  \|\chi_A R_{\La'} \chi_B\|\\
   &=&  \|\chi_A (\ph R_{\La'} - R_\La \ph)\chi_B\|
   \qquad (\ph|_A=1, \ph|_B=0 )\\
   &\stackrel{\text{Lemma } \ref{s006}}{=}&
   \|\chi_A (\ph R_\La (D\ph'+ \ph'D)   R_{\La'} \chi_B\|  \\
   &\leq& \underbrace{\|\chi_A \ph R_\La D\ph'  R_{\La'} \chi_B\|}_{(\ast)}
   + \underbrace{\|\chi_A \ph R_\La \ph'D   R_{\La'}
   \chi_B\|}_{(\ast\ast)}.
\end{eqnarray*}
We start with the analysis of $(\ast)$. If $\tilde{\Om} :=
\text{int}\,(\La_{L-2}(x)\setminus \La_{L-10}(x))$ it holds that
\begin{eqnarray*}\label{g034}
    (\ast) &=& \|\chi_A \ph R_\La D \chi_{\tilde{\Om}}\,
    \chi_\Om \ph'  R_{\La'} \chi_B\| \nonumber\\
    &\leq& \|\ph'\|_\infty \, \underbrace{\|\chi_A \ph R_\La D
    \chi_{\tilde{\Om}}\|}_{(\ast\ast\ast)}
    \, \|\chi_\Om   R_{\La'}  \chi_B\|.
\end{eqnarray*}
The term $(\ast\!\ast\!\ast)$ can be now controlled with the help
of Lemma~\ref{l006}. We put
\begin{equation*}
    f\in L_2(\Ga\cap\La),\qquad g:=\chi_A f, \qquad u:=R_\La g.
\end{equation*}
Then $u\in D(\foh)$ and
\begin{equation*}
    (\foh_\La-E) (u,w) = (g \sk w)
\end{equation*}
for all $w\in D(\foh_\La)$. Furthermore, we have $g|_\Om=0$ as
well as dist\,$(\partial\Om,\partial\tilde{\Om})=1$. Consequently,
Lemma \ref{l006} is applicable and it gives
\begin{eqnarray*}
   \| \chi_{\tilde{\Om}} u' \| &\leq& C_1(q_-,q_+, I)\, \|u\|_{L_2(\Ga\cap\Om)} \\
   &=& C_1(q_-,q_+, I)\, \|\chi_\Om R_\La \chi_A f\|,
\end{eqnarray*}
i.e.
\begin{equation*}\label{g036}
    (\ast\!\ast\!\ast) \leq C_1\,(q_-,q_+, I)\,\|\chi_\La^\text{out} R_\La \chi_A \|.
\end{equation*}
The term $(\ast\!\ast)$ can be treated in a similar way.
$\;\square$

\subsection{The Wegner estimate}\label{c4}

The Wegner estimate represents a statement about the probability
that the operator $H_\La^N(\om)$, restricted to a $\Ga-$ebdd. box
$\La=\La_l(x)$ centered at $x\in\IZ^d$, will have  an eigenvalue
near some fixed energy. Typically -- and sufficiently for our
multiscale analysis -- this probability is polynomially bounded in
terms of the box volume.
\begin{satz}[Wegner estimate]\label{s002}
   For each $R>0$ there exists a constant $C_{R}$ such that for all
   $\Ga-$ebdd. boxes $\Lambda=\Lambda_{l}(i),\, i\in \IZ^{d},$
   and all intervals  $I \subset (-R,R)$ of length $|I|$ the
   following estimate holds:
    \begin{equation*}\label{g020}
      \IP\{\sigma(H_{\La}^N(\omega))\cap I\neq
      \emptyset\}\,\leq\, C_{R}\cdot|\Lambda|^{2}\cdot|I|^{\alpha}.
    \end{equation*}
\end{satz}
Before we start with the proof let us recall the following
elementary lemma from \cite{Sto00}.

\begin{lemma}\label{l002}
   Let $J$ be a finite index set, $\mu$
   a H\"older continuous probability measure on $\IR^d$
   with H\"older exponent $\al$,
   $\mu^{J}:=\otimes_{i\in J}\mu$ the product measure on $\IR^J$.
   Let $\Phi: \IR^J\Nach\IR$ a  monotone function, for which there
   are constants $\de$ and $a>0$ s.t. for all $t\in [0, \de],
   \,q\in\IR^J$ we have
    \begin{equation}\label{g016}
      \Phi(q+t(1,\ldots,1))-\Phi(q)\,\geq \, t\cdot a.
    \end{equation}
   Then for each interval $I$ of length smaller than
   $\ep\leq a\delta$ the following estimate holds:
   \begin{equation*}\label{g018}
      \mu^{J}(\{q: \Phi(q)\in I\})\leq |J|\cdot
      \left(\frac{\ep}{a}\right)^{\al}.
   \end{equation*}
\end{lemma}
\emph{Proof of Theorem \ref{s002}: } We start with an estimate for
the number of eigenvalues smaller than a given energy $R$. To this
aim we define the Neumann-decoupled operator
$-\De_\La^\text{N,dec}$ via its quadratic form
\begin{eqnarray*}\label{g014}
  D(\foh_\La^\text{N, dec})&=&  \oplus_{e\in E(\Ga\cap\La)} W_2^1(0,1), \\
  \foh_\La^\text{N, dec}(f,g)&:=&
  \sum_{e\in E(\Ga\cap\La)} (f'\sk g')_{L_2(\Ga\cap\La)}.
\end{eqnarray*}
By a direct calculation the eigenvalues of this operator are
$\frac{\pi^2}{4}n^2,\,n\in \IN_0,$ with the multiplicity
$\sharp\{E(\Ga\cap\La)\}\leq d\cdot l^d = d|\La|$. Hence there
exists a constant $\tilde{C}_R$ s.t. for the n-th eigenvalue,
counting multiplicity, it holds that
\begin{equation*}
    E_n(-\Delta_\La^{\text{N, dec}})>R \qquad \text{for }
    n>\tilde{C}_R|\La|.
\end{equation*}
Now we have
\begin{equation*}
    H_{\Lambda}^N(\omega) \geq (H_0 + q_-)_\La^N\geq
    -\Delta_\La^{\text{N, dec}}
\end{equation*}
since $q_-\ge 0$ by assumption, and thus by min-max principle the
corresponding inequality for the $n$-th eigenvalues. Using the
previous inequality we get
\begin{equation}\label{g022}
    \IP \{\sigma(H_{\Lambda}^N(\omega))\cap I\neq\emptyset\}
    \leq \sum_{n\leq \tilde{C}_{R}\cdot |\Lambda|}
    \IP\{E_{n}(H_{\Lambda}^N(\omega))\in
    I\}.
\end{equation}
Next we estimate the terms of the sum by means of Lemma
\ref{l002}. Because of the independence of $H_\La(\om)$ of
coupling constants outside $\La$ we have
\begin{eqnarray*}
  \IP\{E_{n}(H_{\Lambda}(\omega))\in  I\} &=& \mu^{E(\Ga)}
  \{\om \sk E_{n}(H_{\Lambda}(\omega))\in  I\}  \\
   &=& \mu^{E(\Ga\cap\La)} \{\tilde{\om}=(\om_e)_{e\in E(\Ga\cap\La)}
   \sk  E_{n}(H_{\Lambda}(\omega))\in  I
   \}.
\end{eqnarray*}
By $\Phi(\tilde{\om}):= E_{n}(H_{\Lambda}(\tilde{\omega})) =
E_{n}(H_{\Lambda}(\omega))$ a monotone function on
$\IR^E(\Ga\cap\La)$ is defined, and it fulfills condition
(\ref{g016}) because
\begin{eqnarray*}
   H_{\Lambda}(\tilde{\om}+t(1,\ldots,1))&=&
   -\De+ \sum_{e\in E(\Ga\cap\La)} (\om_e+t)\chi_e  \\
   &=& H_{\Lambda}(\tilde{\om})+t.
\end{eqnarray*}
Hence by  Lemma \ref{l002} we have
\begin{eqnarray*}
  \IP\{E_{n}(H_{\Lambda}(\omega))\in   I\} &\leq& \sharp E(\Ga\cap\La)\cdot |I|^\alpha \\
   &\leq & d|\,\La|\,|I|^\alpha,
\end{eqnarray*}
which in combination with (\ref{g022}) yields the assertion.
$\;\square$

\subsection{Expansion in generalized eigenfunctions}\label{c8}

Now we come to the last statement needed for the multiscale
analysis, namely that polynomially
bounded generalized eigenfunctions exist spectrally a.s.

We want to use the main result from \cite{BMSt03}, that gives the
polynomial boundedness in terms of the intrinsic metric (see
\cite{Stu94}) generated by the free Laplacian $H_0$ on the graph.
Using the embedding of our graph into $\IR^d$ it can easily be
seen that the intrinsic metric is equivalent to the Euclidean one
on $\IR^d$, and consequently, after adjusting some constants the
statement can be written in terms of absolute values as well. We
start by checking the assumptions of \cite{BMSt03}. First of all
one has to show that the form $\foh_0$ associated with the free
Laplacian is a Dirichlet form. Note that $\|\cdot\|_{\foh_0}$ is
equivalent to the norm $\|\cdot\|_{W_2^1(\Ga)}$ so $\foh_0$ is
closed. For $u\in D(\foh_0)$ which is real-valued we have $|u|\in
D(\foh_0)$, and therefore
\begin{equation*}
    \foh_0(|u|) = \sum_{e\in E(\Ga)} \int_0^1 (\text{sgn}\, u_e(x) u_e'(x))^2
    \,dx \,=\, \foh_0(u).
\end{equation*}
If $u$ is in addition nonnegative, we have $u\wedge 1 \in
D(\foh_0)$ and
\begin{equation*}
   \foh_0(u\wedge 1) = \sum_{e\in E(\Ga)} \int_0^1 u_e'(x)^2
   \cdot 1_{[u_e<1]}(x)\, dx
   \,\leq\, \foh_0(u).
\end{equation*}
Obviously $\foh_0$ is strongly local and regular -- see, e.g.,
\cite{BMSt03} for definitions.

The next point is that the volume of balls with respect to the
intrinsic metric $\rh$ does not grow too fast as $R \Nach \infty$.
Because the graph is  embedded into $\IR^d$ and the intrinsic
metric and the $\|\cdot\|_1$-metric are equivalent,  the volume of
the ball $B_R^\rh(x)$ can be estimated by the number of edges
contained inside a box  $\La_{2R}(x)$ and hence by $c_d R^d$ for
large $R$.

Finally, the third assumption to be checked is that $e^{-tH_0}$ is
bounded as a map from $L_2(\Ga)$ to $L_\infty(\Ga)$ for some
$t>0$. To this aim we employ the following extension of the
ultracontractivity result \cite{KMS06}, Lemma 3.2, demonstrated by
using the same method as in the cited paper.

\begin{lemma}\label{l014}
For $t\in (0,1]$ it  holds that
\begin{equation*}
    \|e^{-tH_0}\|_{L_2(\Ga)\nach L_\infty(\Ga)} \leq
    c t^{-\frac{1}{4}}.
\end{equation*}
\end{lemma}
\textbf{Proof:} By \cite{Ou05}, Thm. 6.3 ff, see also \cite{Na58,
FSt86, Dav89}, it is sufficient to show that
\begin{equation*}
    \|f\|_{L_2(\Ga)}\leq C\cdot \|f\|_\foh^{\frac{1}{3}}
    \cdot\|f\|_{L_1(\Ga)}^{\frac{2}{3}}
\end{equation*}
for $f\in D(\foh)\cap L_1(\Ga)$. Now, by \cite{Ga59, Ni59}, or by
\cite[Sect.~1.4.8]{Ma85}, we have the following Nash type
inequality for  $u\in W_2^1(0,1)$:
\begin{eqnarray*}
   \|u\|_{L_2(0,1)} &\leq& c_1 \cdot \big(\|u'\|_{L_2(0,1)}
   +\|u\|_{L_1(0,1)} \big)^\frac{1}{3}\cdot
   \|u\|_{L_1(0,1)}^\frac{2}{3} \\
   &\leq& c_1 \cdot \|u\|_{W_2^1(0,1)}^\frac{1}{3} \cdot
   \|u\|_{L_1(0,1)}^\frac{2}{3},
\end{eqnarray*}
where in the second step the H\"older inequality has been applied
to $u\cdot 1$. For $f\in D(\foh)\cap L_1(\Ga)$ we have by another
application of H\"older inequality
\begin{eqnarray*}
    \|f\|_{L_2(\Ga)}^2 &=& \sum_{e\in E(\Ga)} \|f_e\|^2_{L_2(0,1)}   \\
    &\leq& c_1^2 \! \sum_{e\in E(\Ga)}  \|f_e\|_{W_2^1(0,1)}^\frac{2}{3} \cdot
   \|f_e\|_{L_1(0,1)}^\frac{4}{3} \\
    &\leq& c_1^2 \,(\sum_{e\in E(\Ga)}  \|f_e\|^2_{W_2^1(0,1)})^\frac{1}{3}
    \cdot
    (\sum_{e\in E(\Ga)} \|f_e\|_{L_1(0,1)} )^\frac{4}{3}  \\
    &=& c_2\cdot \|f\|^\frac{2}{3}_\foh \cdot \|f\|_{L_1(\Ga)}^\frac{4}{3}.
\end{eqnarray*}
\medskip

With these assumptions, given using the arguments in the opening
of the section, \cite{BMSt03} yields the following result:
\begin{satz}\label{s016}
For spectrally a.a. $E\in \sigma(H)$ there exists a generalized
eigenfunction  $\ph$
\begin{equation*}
   (1+ |\cdot|^2)^{-\frac{m}{2}} \ph\in L_2(\Ga).
\end{equation*}
satisfying for any $m > \frac{d+1}{2}$.
\end{satz}
This completes the necessary input for the use of Theorem 3.2.2
from \cite{Sto01} and thus the proof of Theorems 2.1 and 2.2.

\subsection*{Acknowledgments}

The research was partially supported by GAAS and MEYS of the Czech
Republic under projects A100480501 and LC06002 and by the DFG.

\end{document}